\documentclass{tcibook}
\usepackage{fancyhea}
\usepackage{work}
\usepackage{bm}       
\usepackage{graphicx}
\usepackage{multirow}
\usepackage{lineno}
\usepackage{threeparttable}
\usepackage[normalem]{ulem}
\usepackage{color}
\usepackage[colorlinks = true,
            linkcolor = blue,
            urlcolor  = blue,
            citecolor = blue,
            anchorcolor = blue]{hyperref}

\usepackage{amsmath}
\usepackage{xspace}
\usepackage{subfigure}
\usepackage{scalefnt}

\usepackage{sidecap} 
	\sidecaptionvpos{figure}{b} 
\usepackage{caption}
\usepackage{commath}

\def\beq{\begin{equation}}
\def\eeq#1{\label{#1}\end{equation}}
\def\eeqn{\end{equation}}

\newenvironment{Eqnarray}%
   {\arraycolsep 0.14em\begin{eqnarray}}{\end{eqnarray}}
\def\beqa{\begin{Eqnarray}}
\def\eeqa#1{\label{#1}\end{Eqnarray}}
\def\eeqan{\end{Eqnarray}}

\let\bar=\overbar

\def\lsim{\mathrel{\raise.3ex\hbox{$<$\kern-.75em\lower1ex\hbox{$\sim$}}}}
\def\gsim{\mathrel{\raise.3ex\hbox{$>$\kern-.75em\lower1ex\hbox{$\sim$}}}}

\def\del{\partial}
\def\Dslash{\not{\hbox{\kern-4pt $D$}}}
\def\dslash{\not{\hbox{\kern-2pt $\del$}}}
\def\pslash{\not{\hbox{\kern-2pt $p$}}}
\def\ETmiss{\not{\hbox{\kern-4pt $E$}}_T}

\def\Dlr{\mathrel{\raise1.5ex\hbox{$\leftrightarrow$\kern-1em\lower1.5ex\hbox{$D$}}}}

\def\MSB{{\bar{M \kern -2pt S}}}
\def\msb{{\bar{\scriptsize M \kern -1pt S}}}

\def\drb{{\bar{\scriptsize D \kern -1pt R}}}

\begin{document}


\pagenumbering{roman}

\parindent=0pt
\parskip=8pt
\setlength{\evensidemargin}{0pt}
\setlength{\oddsidemargin}{0pt}
\setlength{\marginparsep}{0.0in}
\setlength{\marginparwidth}{0.0in}
\marginparpush=0pt


\pagenumbering{arabic}

\renewcommand{\chapname}{chap:intro_}
\renewcommand{\chapterdir}{.}
\renewcommand{\arraystretch}{1.25}
\addtolength{\arraycolsep}{-3pt}

\newcommand\snowmass{\begin{center}\rule[-0.2in]{\hsize}{0.01in}\\\rule{\hsize}{0.01in}\\
\vskip 0.1in Submitted to the  Proceedings of the US Community Study\\ 
on the Future of Particle Physics (Snowmass 2021)\\ 
\rule{\hsize}{0.01in}\\\rule[+0.2in]{\hsize}{0.01in} \end{center}}

\newcommand{\NNLO}{NNLO\xspace}
\newcommand{\NLO}{NLO\xspace}
\newcommand{\LO}{LO\xspace}
\newcommand{\NcLO}{N$^3$LO\xspace}

\newcommand{\GEANTfour}{{\textsc{Geant4}}\xspace}
\newcommand{\GEANTthree}{{\textsc{Geant3}}\xspace}
\newcommand{\GEANTV}{{\textsc{GeantV}}\xspace}
\newcommand{\VECGEOM}{{\textsc{VecGeom}}\xspace}
\newcommand{\PYTHIA}{{\textsc{pythia}}\xspace}
\newcommand{\PYTHIAeight}{{\PYTHIA\textsc{8}}\xspace}
\newcommand{\MADGRAPH}{\textsc{MadGraph}\xspace}
\newcommand{\MCATNLO}{\textsc{mc@nlo}\xspace}
\newcommand{\MGvATNLO}{\MADGRAPH{}5\_a\MCATNLO}
\newcommand{\MGGPU}{\textsc{MadGraph4GPU}\xspace}
\newcommand{\MADFLOW}{\textsc{MadFlow}\xspace}
\newcommand{\TENSORFLOW}{\textsc{TensorFlow}\xspace}
\newcommand{\BLOCKGEN}{\textsc{BlockGen}\xspace}
\newcommand{\POWHEG}{{\textsc{powheg}}\xspace}
\newcommand{\SHERPA}{{\textsc{sherpa}}\xspace}
\newcommand{\HERWIG}{{\textsc{herwig}}\xspace}
\newcommand{\LHAPDF}{{\textsc{lhapdf}}\xspace}
\newcommand{\HEPMC}{{\textsc{HepMC}}\xspace}
\newcommand{\RIVET}{{\textsc{rivet}}\xspace}
\newcommand{\EVTGEN}{{\textsc{EvtGen}}\xspace}
\newcommand{\TAUOLA}{{\textsc{tauola}}\xspace}
\newcommand{\ALPGEN}{{\textsc{alpgen}}\xspace}
\newcommand{\PHOTOS}{\textsc{photos}\xspace}
\newcommand{\DIRE}{\textsc{dire}\xspace}
\newcommand{\VINCIA}{\textsc{vincia}\xspace}
\newcommand{\MCFM}{\textsc{mcfm}\xspace}
\newcommand{\GENIE}{\textsc{genie}\xspace}
\newcommand{\NEUT}{\textsc{neut}\xspace}
\newcommand{\NUWRO}{\textsc{NuWro}\xspace}
\newcommand{\GIBUU}{\textsc{GiBUU}\xspace}
\newcommand{\ACHILLES}{\textsc{achilles}\xspace}
\newcommand{\MCNNTUNES}{\textsc{mcnntunes}\xspace}
\newcommand{\PROFESSOR}{\textsc{professor}\xspace}
\newcommand{\APPRENTICE}{\textsc{apprentice}\xspace}
\newcommand{\NUISANCE}{\textsc{nuisance}\xspace}
\newcommand{\BEAGLE}{\textsc{BeAGLE}\xspace}
\newcommand{\DiffExp}{\textsc{DiffExp}\xspace}
\newcommand{\AMFlow}{\textsc{AMFlow}\xspace}
\newcommand{\MATRIX}{\textsc{matrix}\xspace}
\newcommand{\NNLOJET}{\textsc{NNLOJet}\xspace}
\newcommand{\MADLOOP}{\textsc{MadLoop}\xspace}
\newcommand{\WHIZARD}{\textsc{whizard}\xspace}
\newcommand{\AMEGIC}{\textsc{amegic}\xspace}
\newcommand{\COMIX}{\textsc{comix}\xspace}
\newcommand{\HELAC}{\textsc{helac}\xspace}
%
%
\newcommand{\PYSECDEC}{\textsc{pySecDec}\xspace}
\newcommand{\FIESTA}{\textsc{fiesta}\xspace}
\newcommand{\HELACNLO}{\textsc{Helac-NLO}\xspace}
\newcommand{\NLOX}{\textsc{nlox}\xspace}
\newcommand{\OPENLOOPS}{\textsc{OpenLoops}\xspace}
\newcommand{\RECOLA}{\textsc{recola}\xspace}
\newcommand{\FIRE}{\textsc{fire}\xspace}
\newcommand{\REDUZE}{\textsc{reduze}\xspace}
\newcommand{\LITERED}{\textsc{LiteRed}\xspace}
\newcommand{\KIRA}{\textsc{kira}\xspace}

%
%
\newcommand{\GRID}{\textsc{Grid}\xspace}
\newcommand{\QUDA}{\textsc{Quda}\xspace}
\newcommand{\Chroma}{\textsc{Chroma}\xspace}
\newcommand{\CPS}{\textsc{CPS}\xspace}
\newcommand{\MILC}{\textsc{MILC}\xspace}

\renewcommand\thesection{\arabic{section}}
\renewcommand\thesubsection{\thesection.\arabic{subsection}}

\title{\vspace{-5\baselineskip}\snowmass CompF2: Theoretical Calculations and Simulation\\ Topical Group Report}
\author{Peter Boyle$^a$, Kevin Pedro$^b$, Ji Qiang$^c$\\
\\
\parbox{\linewidth}{\centering
$^a$ {\normalsize \href{mailto:pboyle@bnl.gov}{pboyle@bnl.gov}; Physics Department, Brookhaven National Laboratory, Upton, NY 11973, USA}\\
$^b$ {\normalsize \href{mailto:pedrok@fnal.gov}{pedrok@fnal.gov}; Scientific Computing Division, Fermi National Accelerator Laboratory, Batavia, IL 60510, USA}\\
$^c$ {\normalsize \href{mailto:jqiang@lbl.gov}{jqiang@lbl.gov}; Accelerator Technology and Applied Physics Division, Lawrence Berkeley National Laboratory, Berkeley, CA 94720, USA}
}
}
\date{September 16, 2022}

\maketitle

\section{Executive Summary}

We discuss the challenges, potential solutions, and needs facing six diverse but related topical areas that span the subject of theoretical calculations and simulation in high energy physics (HEP):
cosmic calculations, particle accelerator modeling, detector simulation, event generators, perturbative calculations, and lattice QCD (quantum chromodynamics).

The challenges arise from the next generations of HEP experiments, which will include more complex instruments, provide larger data volumes, and perform more precise measurements.
Calculations and simulations will need to keep up with these increased requirements.
The other aspect of the challenge is the evolution of computing landscape away from general-purpose computing on CPUs and toward special-purpose accelerators and coprocessors such as GPUs and FPGAs.
These newer devices can provide substantial improvements for certain categories of algorithms, at the expense of more specialized programming and memory and data access patterns.

The adaptation of existing CPU-based software to specialized hardware can provide solutions to the experiment-driven challenges, but this requires additional expertise and substantial effort.
Artificial intelligence (AI) and machine learning (ML) can also provide solutions to some problems that are more immediately amenable to acceleration on coprocessors.
US HEP should, to the extent possible, guide continued development of computing hardware to be maximally useful for solving HEP computing challenges.
Important developments include: faster hardware, availability of general-purpose CPUs, universal programming interfaces and portability, automated memory hierarchy, and early access to new hardware.

Long-term support of careers for software developers and maintenance for common software tools is essential to the future of HEP calculations and simulations.
Efficient utilization of the aforementioned new compute accelerators will require substantial efforts to port existing software to these new paradigms and/or to adopt new AI/ML techniques.
Needs in this area include new organizations to ensure software sustainability, permanent software development staff at national labs, joint lab-university appointments for computationally-focused scientists, and increased training across the field.

Computing in HEP has long struggled with issues of diversity and inclusion.
As the intersection of two highly specialized topics, there is an especially high level of knowledge and skills required to contribute successfully.
To get to this level, individuals usually must seek expertise outside of traditional HEP education and training, which typically does not cover computing topics in detail.
This poses a discouraging burden, especially for those without connections to such expertise or without the resources to take on such additional work.
We hope that substantially expanded support for software and computing training and for software development as a viable career in the HEP ecosystem will alleviate this burden.
A larger, healthier HEP computing community can be more welcoming and accessible to participants from all backgrounds.

\section{Introduction}

The Snowmass Theoretical Calculation and Simulation topical 
working group is cross-cutting and underpins multiple scientific domains.
The computing topics discussed overlap with almost
all aspects of Snowmass science and almost every Frontier. 
Increasingly, computation is central to most theoretical and experimental scientific endeavors.

The great challenges of science, and in particular physical science supported by the Department of Energy and the National Science Foundation,
have a long history of pushing the boundaries of computational science and computing technology. The current
Snowmass study indicates this will continue on many fronts.

The Theoretical Calculation and Simulation group spans a number of sub-topical themes representing the American Physical Society (APS) Division of Particles and Fields (DPF) Snowmass scientific interests:
Cosmic Calculations, Particle Accelerator Modeling, Detector Simulation, Physics (Event) Generators, Perturbative Calculations,
and Lattice QCD (quantum chromodynamics).
This report first introduces the common challenges and the modern computing landscape in Section~\ref{sec:crosscut}, discusses each subtopic in more detail in Sections~\ref{sec:cosmic}--\ref{sec:lattice}, and concludes with recommendations in Section~\ref{sec:concl}.

\section{Cross-cutting computational challenges}\label{sec:crosscut}

The computational landscape of scientific algorithms can broadly be divided into high-performance computing (HPC) and
high-throughput computing (HTC) problems.
HPC requires the coordinated and cooperative processing of a single large computational problem
across multiple nodes with data
being exchanged as part of the calculation, often using high-performance
interconnects.
In contrast, HTC processes a large volume of completely independent computational work units.

Depending on the problem type, both HPC
and HTC may have a high volume of relatively simple 
bulk floating point
arithmetic that can be accelerated 
using highly parallel special-purpose computing hardware.
Some calculation and simulation tasks, and particularly those that process large data arrays or are machine learning-based, may obtain substantial speed increases from these hardware accelerations. In other cases, algorithms that involve substantial serial performance or many
data dependent logical flow decisions will benefit most from hardware that is suited
to efficient (and power efficient) serial execution.
Detector simulation, event generation, and continuum theoretical calculations are typically HTC tasks, while accelerator modeling, cosmic calculations, and lattice gauge theory are more often HPC tasks.
However, exceptions exist, and the use of different algorithms or programming paradigms can sometimes convert HTC to HPC.

Memory bandwidth has presented a fundamental
rate limit to many calculations.
It is common to include computational
accelerators that can provide substantially greater
floating point throughput and memory bandwidth
as coprocessor cards. The most common
of these are graphics processor units (GPUs),
but Field Programmable Gate Arrays (FPGAs) and other technology options are also common.
These often provide distinctly addressable
high performance
memory to help deliver good performance.
Coprocessors are often highly parallel and have limited or
fixed function operations, so they are only usable with
certain classes of algorithms.
Machine learning (ML) algorithms are especially suited to be accelerated on these devices, as described in Section~\ref{subsec:comp}. On the other hand, 
while event reconstruction is easy to parallelize by processing independent
events independently on independent serial processor cores of any form, the standard GPU thread model serializes execution when different threads in a group make different data-dependent logical flow decisions (for example, making different event-specific patterns of subroutine calls).

From an electronics perspective, it is easier to provide multiple types
of memory (small and fast, large and slow) organized in distinct circuit boards to support a high floating point density. 
The memory technologies are currently split into on-chip caches, in-package DRAM or ``high bandwidth memory'', off-package DRAM, and increasingly non-volatile memory. These are listed in order of increasing capacity and decreasing memory bandwidth. The appearance of 2.5D and 3D integration technologies has 
been transformational and this system-level innovation will likely continue to return greater
memory bandwidth over time, giving continued practical gains in high performance computation.
Such hierarchical memory systems will likely remain in the future,
as they are dictated by physics constraints. The degree of hierarchy is
increasing and may grow in the future, independent of the specific processor architecture. 
In this sense, the trends identified are likely reasonable over a ten-year timeframe.

Scientific productivity will be enhanced if data placement is handled automatically and moved infrequently: for example, this should be automatically and efficiently handled by the Unix virtual memory system. This would avoid a requirement that every single scientific programmer laboriously manage the placement and movement of data to specific locations, such as caches.
The managing of distinct pointers to distinct ``host'' and ``device'' memory spaces is both
a recurrent source of programmer error and requires considerable care to ensure that the allocated GPU memory does not exceed the limited capacity, while automatic paging avoids this requirement.
Section~\ref{Sec:software-engineering} discusses more considerations relevant to scientific programming with modern heterogeneous computing.

Interconnects, used for both message passing and filesystem access, are becoming an increasing bottleneck since accelerated computing nodes are becoming rapidly
more powerful while interconnect performance gains have not always matched pace. Technologies
such as silicon photonics, allowing the integration of optical transceivers in standard silicon chips, are
a promising basic technology that could help address interconnect bottlenecks. Similarly, high speed links between GPUs within a computing node are a key enabling technology for many problems.

File systems are a key element of any scientific workflow. Data access should
be sufficiently fast not to present a bottleneck. Modern computing systems
use parallel file systems, such as Lustre, giving up to terabyte/s aggregate access rates to disk. The use of non-volatile memory elements can both cache bursts of data and accelerate metadata access. Data migration to cold
or longer term storage can be carried out through either user input or hierarchical storage managers. It is important that these keep pace with growing data volumes and increasing computation speed.

\subsection{Computing hardware acceleration}\label{subsec:comp}

Fixed-function acceleration
is possible for
problems involving dense matrix multiplication,
with tensor cores in GPUs and other hardware specifically designed
to handle large matrices. These operations are commonly performed using 
64-bit and 32-bit floating point arithmetic, or even 
on 16-bit floating point arithmetic. Mixed precision functional units are common. 
Fixed-function dense matrix
hardware such as tensor cores do not necessarily accelerate all algorithms.
ML is one common and growing source of matrix multiplication operations that can be accelerated in this way.

ML can broadly be decomposed into ``training'' and ``inference'' phases. Training often involves scientist supervision
for parameter tuning, with time to results being critical. 
Distributed ML training parallelizes the training process.
The use of small ``mini-batches''
in stochastic gradient descent leaves the problem very sensitive to 
Ahmdahl's law slowdown. In the massively distributed
limit~\cite{Kurth:2017poy}, the summation across an interconnect of a gradient vector
calculated across many randomly selected training samples in
a minibatch is a significant performance bottleneck~\cite{Baidu}. 

Inference with a previously trained network is typically
performed on many independent events of data and is in many cases a
trivially parallel (HTC) problem. However, when inference is used
in either hard or soft real-time situations (as occurs in HEP experiments),
the latency requirements impose challenging performance issues when traditional CPUs are used.

Accelerators from Graphcore and Cerebras are good examples of
ML-optimized accelerators in
addition to more standard GPUs by Nvidia, AMD, and Intel. 
Reconfigurable computing hardware 
such as FPGAs or spatial accelerators are powerful
options for certain algorithms. 
They may be accessed via a dataflow
model where a hardware circuit 
is configured in a non-von Neumann approach
to stream data from memory, calculate, and store results in memory.
They may be particularly effective in eliminating software latency 
in real-time inference environments~\cite{Harris:2022qtm,Carini:2022vut}.
Algorithms that are I/O-bound or executed in situations requiring reduced power consumption may also be more efficiently accelerated on FPGAs.

\subsection{Software challenges}
\label{Sec:software-engineering}

The computing landscape at this time displays an exciting 
proliferation of competing computer architectures.
It opens vibrant and healthy opportunities for innovation and competition and
for the introduction of new ideas.
However, it also risks introducing special-purpose or
less-general features, compared to previous systems. The memory spaces
are often fragmented with different pointer types and data locations.
There are multiple instruction-set architectures in use, each
with a different programming interface or model.
A single program frequently has different segments that target different
types of instruction sets within a single computer node. Indeed, 
some of the computing models available (such as FPGAs or spatial
architectures) are not von Neumann computers. Others have dedicated
fixed-function acceleration of matrix-matrix multiplication for machine
learning.

This diversity presents a significant challenge to the scientific programmer
with either legacy code or the need to use more than one system.
There are a number of different programming models
one must consider in the present computing landscape. The following can all in principle
be combined with MPI message passing.

Present Exascale and Pre-exascale systems support:
\begin{itemize}
    \item traditional CPU core, SIMD floating point instructions, and OpenMP threading,
    \item CUDA-based GPU programming,
    \item HIP-based GPU programming,
    \item SYCL-based GPU and FPGA programming, and
    \item OpenMP 5.0 offload GPU programming.
\end{itemize}

Future systems may support standards-based 
C++ parallel STL offload programming;
however, multi-system compiler support is currently absent.
The challenges of writing high-performance and portable code are threefold:
 syntactical differences, 
 semantic differences, and
 data placement.
The proliferation of models represents a substantial cost and effort overhead
for science exploitation, since multiple implementations of software must be developed and maintained.
At this time, due to the diversity of models, abstraction and support for multiple
programming models is likely the safest option for portable efficiency on near-term computers.
For projects that wish to reuse interfaces to acceleration,
several good portability options exist: Alpaka and Department of Energy ASCR projects Kokkos and RAJA. Standards-based programming, such as SyCL, OpenMP, or C++ parallel STL, is to be encouraged.

It is important to recognize that
not all algorithms are amenable to map to accelerated hardware.
This must be considered in the HEP computing landscape to avoid leaving important needs unaddressed. 
There is much to be said
for an element of the computing roadmap using general-purpose processor cores combined with advanced memory technologies such
as high bandwidth or hierarchical memory.
In this way, the long
tail of problems with computational challenges that would
be either impossible or prohibitively expensive to port to
accelerated hardware may continue to be addressed.

\section{Cosmic Calculations}\label{sec:cosmic}

In the next decade, a number of powerful observational facilities will
start to operate. These next-generation cosmological observations will provide great opportunities to study the early history of universe, the mysteries of dark matter and dark energy, the fundamental particle physics, and possible modifications of Einstein 
gravity. 
In order to successfully achieve the observational goals, a state-of-the-art simulation and modeling program is needed~\cite{Alvarez:2022}. Cosmological simulations provide an invaluable tool to optimize the observation design, to interpret observation data, and to help unearth the underlying physics. Next-generation cosmological simulations will increasingly focus on physically rich high-fidelity simulations that can directly connect with observational outputs from multiple different surveys.

Several types of cosmological simulations have been employed in the cosmology study.
The first-principles N-body simulations that have no free parameters can accurately
describe dark matter fluctuations from the largest observable scales down to scales
deep into the nonlinear regime. However, the N-body approach does not account for
the physics of the baryonic sector. Additional models such as the halo occupation distribution,
sub-halo abundance matching, or other schemes are added on top of a simulation to reconstruct galaxies.
Hydrodynamical simulations provide a reasonably accurate description of the distribution of
baryons, quantify the effects of baryons on various probes of large-scale structure,
 and provide useful results for the distribution and properties
of galaxies and clusters. However, the robustness of the hydrodynamical simulation results
depends on the choices of subgrid models and the
nature of the cosmic probe in the study. In addition to the aforementioned simulations, 
other simulations such as beyond $\Lambda$CDM simulations that involve the solution
of nonlinear Poisson equation in the modified gravity model and radiative transfer
simulations that model reionization process have also been used in cosmological studies.

Different survey experiments include different observables in the study. Signals from galaxy clustering and lensing are key observables for photometric galaxy surveys. 
Spectroscopic instruments are used to measure redshifts, radial velocities, gas dynamics and chemical compositions of galaxies. 
Signals from Baryonic Acoustic Oscillation (BAO) and
redshift space distortions (RSD) measurements are used to extract useful cosmological information.
Signals from the cosmic microwave background lensing measure the integrated mass between the last scattering
surface and us.
The Lyman-alpha forest signal in the spectra of distant quasars is used to constrain thermal properties of the intergalactic medium (IGM). 
The thermal and kinematic Sunyaev Zeldovich (tSZ/kSZ) signals are used to infer the distribution of gas in the universe. The emission from dusty star
has been used to delens the cosmic microwave background (CMB). X-ray observations are used
to calibrate mass estimates. High-redshift from interferometers can
provide useful information about the thermal and ionization state of the intergalactic medium and can be used to learn about dark matter.

The diverse observational probes present significant challenges in cosmological simulations.
One challenge is the computational cost to include multiple probes with 
high precision. These simulations need cover a large volume
but with sufficient fine resolution for a large number of realizations.
In addition, computationally demanding hydrodynamical simulations
are required to simulate some observables such as the thermal and kinetic Sunyaev-Zel'dovich effects, tSZ and kSZ, to infer the distribution of gas in the universe, or Lyman-$\alpha$, to constrain thermal properties of the intergalactic medium.
In some cases, the modeling of certain observables
involves a very large dynamical range. Semi-numerical
simulations are used when the hydrodynamical simulations
cannot cover large enough volume while reaching small enough halos.
Calibrations and benchmarks between different approaches are
important to interpret new observation data.
Another challenge for cosmological simulations is to 
reproduce observations, for example, correlations
between multi-wavelength observables, in consistent galaxy formation models.
It is necessary to
develop simulation models that can be applied to
gravity-only simulations and account for correlations between neutral and ionized gas, stars,
and dust in galaxies and galaxy clusters.
Also, it is not trivial to simulate massive neutrinos since they
make up a non-negligible fraction of the total energy but can be decoupled
when relativistic and have a free streaming scale.
In the N-body simulations, thermal velocity distributions
need to be accounted for in the structure formation calculation on smaller scales.
Covering both the large volume required by the weak lensing
and maintaining the accuracy at small scales required 
by strong lensing presents another
computational challenge in ray tracing. In the modeling for future analyses
of small scale measurements,
baryonic feedback effects that can change the local matter density
must be included in the cosmological simulations.
Besides advancing in terms of increased resolution,
larger volumes and better treatments of known physics,
additional new physics models need to be included as
the reach of surveys expands.
These include models of modified gravity with
nonlinear Poisson equation for dark energy studies and 
a variety of dark matter models (warm dark matter,
interacting dark matter, self-interacting dark matter, ultralight dark matter,
ultraheavy dark matter, multiple component dark matter~\cite{Banerjee:2022}). 
Dedicated cosmological simulations are needed for those new models.

Next-generation cosmological surveys demand better
understanding, mitigation, and control of systematic uncertainties. To attain these goals, it is important to create realistic ``virtual universes'' from cosmological simulations. Such simulations can be used to solve the
inverse problem to infer the physical parameters of different models from the sky survey observation data.
However, it is not computationally practical to model all relevant processes from a first-principles approach
given the vast dynamic range in cosmology.
Fast surrogate
models can be used as effective emulators to speed up the solution of this inverse problem. 
The predictions for cosmological surveys from emulators have been
widely used in cosmological studies. It is desirable that 
these emulators can connect directly to the 
survey observables to extract cosmological parameters. 
With the advance of multi-fidelity hydrodynamical simulations,
it is also desirable to use emulators to optimize 
subgrid model parameters and to gain better
understanding of the interplay between the subgrid model
and cosmology parameters.

Cosmological simulations are computationally intensive and demand the use of state-of-the-art computer hardware. With the arrival of exascale supercomputers, cosmological simulations will need to use computer accelerators efficiently and memory access patterns effectively, as well as performance portable programming models and scalable algorithms. 
Progress has been made to address some of these challenges, for example under the Exascale Computing Project (ECP)
led by the Office of Advanced Scientific
Computing Research (ASCR) in collaboration with other DOE science program offices~\cite{Habib2016,Sexton2021}.
Continuous development and support are extremely important to enable full use of these hardware 
systems. The successful use of the exascale supercomputers will enable gravity-only simulations with 
unprecedented volume coverage and resolution and hydrodynamics simulations with exceptionally 
detailed baryonic physics modeling for the cosmological surveys.
Scalable analysis approaches are as important as the development of simulation codes since 
handling and processing
of very large data sets generated by the simulation codes require large computing resources in their own right.
Besides the development of the simulation codes, the development of analysis tools faces the same challenges as the simulation codes with respect to the efficient usage of the available computer hardware. Co-development of simulation codes and analysis tools is needed for a successful cosmological simulation program.  

Cosmological simulations are critical for the success of
next-generation cosmological observations. The accuracy for
these simulations needs to be carefully checked through 
verification by benchmark among codes and validation with direct comparison against observational data. 
This requires a concerted effort between different code and analysis development teams and
cosmic observation teams. The detailed
connection between the simulations and the survey observations has to be established for a successful validation program.

In summary, the cosmological simulation community 
recommended the following advances in order to meet the above challenges and needs~\cite{Alvarez:2022}.
\begin{itemize}
\item{Development of scientifically rich and broadly-scoped simulations, which capture the relevant physics and correlations between probes}
\item{Accurate translation of simulation results into realistic image or spectral data to be directly compared with observations}
\item{Improved emulators and/or data-driven methods serving as surrogates for expensive simulations, constructed from a finite set of full-physics simulations}
\item{Detailed and transparent verification and validation programs for both simulations and analysis tools}
\end{itemize}

Another area of cosmic calculations is the study of numerical relativity for next-generation gravitational-wave probes of fundamental physics~\cite{Foucart:2022}. The next-generation gravitational-wave detectors will provide great opportunities to study the nuclear physics of
dense matter from the signals of neutron-star or black hole mergers, the nonlinear dynamics of warped spacetime, and the physics beyond general relativity from the signals of binary black holes. To attain these goals requires a new generation of numerical-relativity codes with dramatically improved accuracy that can make effective use of the exascale supercomputers. The higher computational accuracy also results in a significant increase of needs for computational resources in comparison to the present typical weeks to months of runtime on tens to thousands of computer cores. This is due to the higher simulation resolution and the
longer simulation to cover the detectors' sensitive 
frequency spaces.
In addition, many simulations will be necessary to span the 
parameter space of potential signals from black holes and
neutron stars.

A further topic in cosmic calculations is the modeling of neutrino transport in neutrino-driven core-collapse supernovae~\cite{Mezzacappa:2020}. A kinetic description of these neutrinos is needed in this model. The solution for the neutrino transport equation is computationally intensive, involving high-dimensional linear algebra. These simulations have been ported to GPU-based supercomputers and will require more computing resources to improve the resolution of the calculations and to explore different physical scenarios. 

\section{Particle Accelerator Modeling}\label{sec:accel}

Particle accelerators are regarded as one of the most important inventions of the $20^{\mathrm{th}}$ century. They provide an
invaluable tool in high energy physics study. However, particle accelerators are large and complicated scientific 
devices that can
be thousands of meters long (e.g., the LHC's nearly 27 kilometer circumference) with hundreds of thousands of elements.
Their design, construction, and operation are both sophisticated and expensive, and depend critically on computer modeling

In the particle accelerator modeling area, there are 26 related letters of intent submitted to this 
Snowmass process. These letters of intent were summarized in a community-wide white paper that addresses the challenges
and needs in modeling particle accelerators~\cite{Biedron:2022}. Next-generation
high energy particle accelerators and colliders will be more complicated and expensive than the existing accelerators. 
These accelerators may employ both conventional RF cavity and high gradient wakefields from either a plasma channel or a structure waveguide to accelerate charged particle beams
to high energy and use strong superconducting magnets and sophisticated control systems to confine
these beams. Some potential next-generation particle accelerators include very high energy proton-proton colliders, multi-TeV electron-positron colliders, muon colliders, other very high energy machines such as
gamma-gamma and high energy electron-ion colliders, and high
intensity accelerators for neutrino physics study.

To meet the needs of these next-generation accelerators, it is important for particle accelerator modeling tools to
include all types of accelerating structures (e.g., RF-based,
plasma-based, structured-based wakefield, plasmonic) and different accelerator components (e.g. superconducting magnets, structured plasmas), and to target the accelerator
and beam physics thrust grand challenges on intensity, quality, control, safety, and prediction.
For RF-based accelerators, this requires modeling of collective effects with high fidelity over long time scales, 
improving computational speed to allow for statistical ensembles and design optimization including collective effects, 
and improved integration with realistic magnet and RF modeling.
For plasma-based accelerators, this requires supporting
the development of modeling tools and methods that will enable start-to-end simulations that
predict the full six-dimension (+ spin) evolution of beams in a plasma-based linear collider, from their
creation to their final focusing at the interaction point, and include all the intermediate
phases of acceleration, transport, manipulations, and collisions.
For structure-based accelerators, this requires the
development of efficient and accurate algorithms capable of modeling the beam interaction
with its wakefield over long interaction lengths in structures with arbitrary geometries
and constitutive parameters. 
For plasmonic accelerators, this requires the development of new approaches such as
a quantum-kinetic method that incorporates the effects of localized petavolts per
meter plasmonic fields on the underlying quantum mechanical processes such as the
ionic lattice and the energy band structure and to incorporate the multi-physics nature of the
unprecedented strongly electrostatic plasmonic modes.  
For the structured plasma accelerator component, this requires the development and integration
of fluid and kinetic codes to meet the modeling demands for a new class of structured
plasma devices that couple macroscopic plasma properties with strict requirements on kinetic
interactions. The quest for higher accelerating gradients in next-generation particle accelerators
puts increasing demands on better understanding the materials of the accelerating structure and
relevant phenomena such as RF breakdown. This needs the development of automated scale-bridging methods that
can autonomously parameterize higher-level models from large numbers of lower-scale first-principles calculations
to enable predictive materials studies over a broad range of materials and conditions.
Strong magnets will be highly demanded in the next-generation particle accelerators to reduce the machine
size and cost for transverse focusing. Some advanced modeling tools are currently
utilized in the US Magnet Development Program. Novel modeling tools with mixed finite element
method are needed to improve design time, cost, and performance of future superconducting
accelerator magnets.

Next-generation particle accelerators will occupy a large land space and consist of many
different accelerator segments starting from charged particle 
sources and ending with final interaction colliders or targets. 
Different segments of the accelerator are coupled with each other through the charged
particle beam transporting through the entire accelerator
system. The final particle beam quality does not depends on the performance of a few elements, but on the performance
of the entire accelerator system. Without the modeling of the whole accelerator, no project can proceed to 
the building stage. 
In order to evaluate the particle accelerator performance and to attain a global optimal design, end-to-end simulations are needed to 
establish a virtual particle accelerator. The end-to-end simulation denotes a simulation starting with the source that generates charged particles and ending with either the final target or the interaction region. These simulations should include all pertinent physical effects from both first-principles models for high precision accelerator design and fast machine learning-based surrogate models for online particle accelerator tuning. Such a virtual particle accelerator can be used as an emulator of the real physical accelerator in the accelerator operation. One challenge in the end-to-end simulation is to exchange particle distribution
data from different components of the particle accelerator. To solve this problem requires supporting 
efforts to establish standards that would ease the sharing of information
and data across codes as well as standards for interfacing codes~\cite{openPMDstandard}.

Another challenge associated with the high fidelity end-to-end accelerator simulation is that it is extremely time consuming
(especially with the inclusion of a variety of collective effects), varying from days to weeks and months.
The cutting-edge and emerging computing techniques 
including advanced algorithms, AI/ML methods, and quantum computing may substantially 
reduce the computational time of accelerator simulation and enable a fast real-time optimization with end-to-end simulations. 
Advanced algorithms play an important role in the accelerator modeling.
High-fidelity self-consistent simulations cannot be done within a 
reasonable return time without using these algorithms~\cite{Qiang:2006,Qiang:2009}.
It has also been shown that the advanced algorithms can lead to several orders of magnitude
improvement on the computational speed~\cite{Vay:2007}. 
Research and development in this area is needed 
to explore new algorithms that exhibit better properties, to refine the understanding of the 
properties and bottlenecks of existing algorithms, and to remove the bottlenecks on cutting-edge heterogeneous computing hardware (e.g., GPU, FPGA),
and improve the speed and accuracy of accelerator modeling.
Differentiable simulations show great 
potential to reduce the number of simulations needed for optimization and design of particle accelerators.
Using AI/ML surrogate models of the physics simulation can save orders of magnitude computing time~\cite{Edelen:2020}.
A single differentiable simulation can generate the sensitivity information of the final beam
quality with respect to thousands of machine parameters. 
Support is needed for the development of AI/ML modeling techniques and their
integration into accelerator modeling and control systems, with an emphasis on fast executing
(up to real-time) and differentiable models, continual and adaptive learning
for time-varying systems, uncertainty quantification to assess
confidence of model predictions, and physics-informed methods to enable broader model
generalization to new application conditions.
Quantum computing is another area of emerging technology that has the potential to 
transform some areas particle accelerator simulations by providing exponential
improvement of computational efficiency.
Support is needed for quantum computing algorithm
and code development for accelerator modeling, feasibility study on quantum
computing implementation in accelerator modeling. Quantum computing education in
the accelerator community is also needed.

Particle accelerator modeling is computationally intensive and has 
utilized supercomputers for many years. 
It has used both high-performance computing to model collective effects and
high-throughput computing to model single particle dynamics.
With the arrival of exascale supercomputers and new special purpose computers with hardware accelerators (e.g. GPUs), it
is a challenge
to achieve the supercomputer peak performance for modeling
collective effects in particle accelerators due to deep levels of independent memory caches, on top of global and neighboring communications
that are used
to exchange information across multiple computer nodes during the simulation.
As part of this, the workload needs to be uniformly
distributed among many compute units in order to achieve a good parallel efficiency. 
The current and next-generation supercomputers extensively use the computational accelerators
such as GPUs and FPGAs to improve the computational speed of applications. 
Applications related to single particle dynamics can be effectively simulated
on GPUs. It is a challenge for applications including multiple physics models and 
collective effects to use GPUs effectively.
Support is needed to adopt portable, state-of-the-art
programming paradigms and frameworks that support both multi-level parallelization and effective
dynamic load balancing over all levels of compute parallelism, so that simulations run efficiently on the latest computer architectures
with scalable I/O, post-processing
and in situ data analysis solutions to support accelerator machine design and operation.

Particle accelerator modeling codes were developed with the support of a variety of projects
to simulate RF structures, magnets, beam dynamics inside
the accelerator structure, laser and beam interaction with plasma, and beam beam and beam wave interactions. 
These simulation tools are critical for the design and operation of next-generation
high energy particle accelerators. 
Sustainability, reliability, user support and training
of these complicated codes are potential challenges in the accelerator community.
Further support is needed to maintain these
codes with documentation, testing, and benchmark examples.
Training classes are needed at the U.S. Particle Accelerator School or other accelerator schools
or workshops to help improve the usage of these codes. 
Another challenge for these accelerator simulation codes is validation with experimental
measurements. This is critical for establishing the confidence in the use of these simulation tools
for next-generation particle accelerator design applications.
The direct comparison between code predictions and accelerator performance and output
may be limited by a number of factors, such as not fully characterized machine settings,
unknown initial phase space distribution, insufficient range or resolution of diagnostics,
and incomplete or missing physics in simulations.
Support is needed to connect simulations with
direct experimental diagnostics and to carry out detailed
benchmark studies with experimental measurements.

The particle accelerator modeling codes include a variety of physics models for different accelerator components
and involve different research groups.
In order to attain the end-to-end simulation for next-generation particle accelerators, different component models
need to be integrated into a workflow and ported onto the latest exascale type computer hardware with GPUs.
New physical models and capabilities will be added in some of these codes to account for new accelerating and focusing methods.
It is desirable to organize the beam and accelerator modeling tools and community through the development of ecosystems of
codes, modular libraries, and frameworks that are interoperable via open community data standards for
both loosely integrated and tightly integrated workflows. 
It is desirable to establish open access data repositories for reuse 
and community-wide AI/ML surrogate model training.
It is also desirable to have an organized particle accelerator modeling
community (e.g. in the form of centers and consortia) including both academic institutes and industry. 
This will improve the efficiency of accelerator modeling for next-generation high energy accelerator applications.

In summary, to meet the above challenges and needs, the following high-level recommendations are provided from the particle accelerator modeling community~\cite{Biedron:2022}:
\begin{itemize}
    \item Develop a comprehensive portfolio of particle accelerator and beam physics modeling tools in support of achieving Accelerator and Beam Physics Thrust Grand Challenges on intensity, quality, control, and prediction.
    \item Develop software infrastructure to enable end-to-end virtual accelerator modeling and corresponding virtual twins of particle accelerators. 
    \item Develop advanced algorithms and methods including AI/ML modalities and quantum computing technologies.
    \item Develop efficient and scalable software frameworks and associated tools to effectively leverage next-generation high-performance and high-throughput computing hardware.
    \item Develop sustainable and reliable code maintenance practices, community benchmarking capabilities, and training opportunities to foster the cooperative application of accelerator software.
    \item Organize the beam and accelerator modeling community in open efforts to (a) foster the development of ecosystems of interoperable codes and tools, libraries, and frameworks, based on API and data standards, (b) establish open access and data workflow repositories, and (c) have an organized governance structure.
\end{itemize}

\section{Detector Simulation}\label{sec:detsim}

Accurate detector simulation is crucial for tasks ranging from detector design to data analysis.
\GEANTfour~\cite{Agostinelli:2002hh,Allison:2006ve,Allison:2016lfl} is the primary tool used for detector simulation throughout HEP~\cite{Banerjee:2022jgv}.
This C++-based software package was originally released in 1998 as a complete rewrite of the earlier FORTRAN-based \GEANTthree~\cite{Brun:1987ma}.
The \GEANTfour collaboration and the HEP community engage in continuous research and development to improve the computational performance, refine the models of physical interactions, and incorporate new technical features and models.
\GEANTfour is the center of an ecosystem that includes numerous software packages serving different needs:
\begin{itemize}
\item Geometry/material modeling and navigation: \VECGEOM, CADMesh, DD4hep, GDML;
\item Physics models: Noble Element Simulation Technique (NEST) for excitation, ionization, etc.; G4CMP for phonons and other condensed matter phenomena; Opticks for optical photon simulation;
\item Other simulation packages (including radiation modeling): MARS, FLUKA, MCNP, SRIM;
\item R\&D projects: G4HepEM, AdePT, Celeritas;
\item Fast simulation engines: DELPHES, TrackToy, bespoke experiment-specific solutions, emerging ML approaches.
\end{itemize}
Some of these packages are developed in very close coordination with \GEANTfour, while others are more independent, using \GEANTfour as a reference or being used as cross-checks for \GEANTfour results.
In addition, \GEANTfour interfaces with event generators (see Section~\ref{sec:evtgen}) such as \PYTHIAeight for decays of long-lived, metastable, or late-forming particles, and with ROOT and CLHEP for various data analysis and mathematical functions.
Custom simulations of electronics responses and other instrumental effects are developed and applied downstream from detector simulation~\cite{Nachman:2022wqs}.

\GEANTfour is one of the most successful examples of community-supported and -maintained software in HEP.
It is even used beyond HEP for medical and astronomical applications.
However, HEP needs are the primary drivers of R\&D on the software, especially improvements to its computing performance.
In particular, the next generation of collider experiments, including the HL-LHC and the various future colliders proposed during the Snowmass process, will pose extreme computing challenges for detector simulation.
Compared to present-day experiments, data volumes will increase by at least an order of magnitude, requiring a corresponding increase in the number of simulated events.
Detectors will increase in complexity and precision, requiring more detailed geometries and more precise physics models that will use even more computation.
At the same time, other aspects of the event processing chain, such as reconstruction, will consume a growing proportion of available computing because of massive increases in simultaneous bunch crossings, or pileup, especially at hadron colliders.
Therefore, \GEANTfour will need to simulate an order of magnitude more events with higher precision while using a smaller proportion of HEP computing.
To ground this prediction in reality, \GEANTfour used 40\% of all computing during the LHC Run 2 for CMS and ATLAS~\cite{HEPSoftwareFoundation:2018fmg}, and ${\sim}70\%$ for LHCb~\cite{Muller:2018vny,Bozzi:2802074}.
For the former, it is projected that only 14--20\% will be available for simulation at the HL-LHC~\cite{CMSOfflineComputingResults,Collaboration:2802918}.

In addition to high energy colliders, experiments of various sizes that search for light or weakly interacting particles rely on \GEANTfour~\cite{FASER:2022yqp,Kahn:2022kae,Roberts:2022ezy}.
Such experiments include DUNE, FASER, MicroBOONE, Muon g-2, Mu2e, COHERENT, and direct dark matter detection experiments DAMIC, DarkSide, DARWIN, DEAP, LZ, NEWS-G, PandaX, PICO, SBC, SENSEI, SuperCDMS, and XENON.
While many of these experimental collaborations and apparatuses are smaller than those at the CERN LHC, they face similar computational challenges to process petabytes of data, including the corresponding simulation of background and signal processes.
The need to meet these challenges with substantially fewer dedicated personnel makes the availability of common software even more important, including both \GEANTfour and packages like NEST and Opticks that provide specialized physical interaction models.
Future neutrino, dark matter, and other similar experiments will obtain even larger data volumes and probe even more precise physics, increasing the computational challenges and the associated need for high-performing and well-supported software.

The increasing breadth and precision of next-generation HEP experiments will require corresponding improvements in the physics models used in \GEANTfour and related software~\cite{Banerjee:2022jgv}.
Some areas of interest include liquid argon signal induction, scintillation materials, Cerenkov light propagation, condensed-matter effects, low-energy responses, and rare background processes and interactions.
In particular, hadronic interaction models are an area where agreement with data could be improved and access to more measurements, such as hadron-argon interactions, could lead to even further improvement.
In general, support for increased effort in these areas is needed to keep up with the demands of new experiments for more precise models, as well as to improve the computational efficiency of the model implementations.

The first major evolution in the \GEANTfour computing model was the introduction of multithreading to reduce memory usage by sharing read-only common data during otherwise embarrassingly parallel event processing~\cite{HEPSoftwareFoundation:2020daq}.
This approach has been adopted by the LHC experiments and Mu2e as the experiment software frameworks have similarly evolved to incorporate multithreading in other event processing steps~\cite{Jones:2022ycw}.
A recently concluded R\&D project called \GEANTV introduced sub-event parallelism using vectorized instructions, primarily targeting many-core CPU architectures, in a new prototype transport engine.
The primary findings of this project were that the achievable speedup compared to \GEANTfour was limited to a factor $2\pm0.5$ and that a small percentage of the speedup actually arose from vectorization~\cite{Amadio:2020ink}.
Nevertheless, the project produced several useful developments, such as \VECGEOM, and lessons, such as the importance of instruction cache locality.
\VECGEOM provides improved code and vectorization for geometry modeling, enabling a ${\sim}15\%$ speedup when used with \GEANTfour.
As an illustration of the level of complexity of \GEANTfour, it required 30 FTE-years, spread over a 5-year period, to re-implement its numerous components, even partially or as prototypes, in \GEANTV: particle transport, geometry modeling and navigation, magnetic field propagation, physical interaction models, and interfaces for experiment software frameworks.

The future of detector simulation computing is turning toward the use of GPUs, targeting in particular HPC facilities that are intended to provide a larger fraction of HEP computing for next-generation experiments.
Because detector simulation is naturally an HTC problem, nontrivial effort is required to adapt the necessary computations to use computer accelerators effectively.
Two ongoing projects, G4HepEM-AdePT and Celeritas, have produced viable prototypes of high-performing GPU-native simulation engines.
In particular, the Celeritas effort is led by US national laboratory personnel and was inspired by the earlier porting of the Shift MC radiation transport engine to GPUs~\cite{Tognini:2022nmd}.
Shift is a less general application than \GEANTfour, focusing on neutral particles with few secondary particles produced, limited physical interactions, simple geometries, and no magnetic field propagation.
Despite the additional complications encountered in addressing the broad scope of HEP detector simulation, the prototypes have demonstrated promising results, achieving a factor of 40 speedup using 1 Nvidia V100 GPU compared to \GEANTfour using 7 Power9 CPUs on Summit~\cite{GPUCommunityMeeting}.
This can be rephrased as a factor of 200 GPU:CPU equivalence comparing a V100 to a higher-end Xeon CPU, also taking into account power consumption and other relevant factors.
Variations in performance from different implementation choices were relatively small, a factor of 2 at most.
Scaling this up to the whole Summit supercomputer suggests that its 27\,648 GPUs can correspond to 5.5 million CPU cores, an order of magnitude larger than the current worldwide LHC computing grid (WLCG) at 500\,000 cores~\cite{WLCG}.
If this performance persists, it has the potential to resolve the major simulation computing challenges.
However, we must not underestimate the increased burden of developing and maintaining coprocessor-friendly code, including aspects such as portability to non-Nvidia GPUs and even future non-GPU architectures.
It should also be noted that the ongoing projects currently implement only some components of \GEANTfour, including models of electromagnetic interactions, but not yet hadronic interactions, for example.
If a complete parallelization of all components is not ultimately possible, the observed speedup will be limited by the remaining serial components, according to Amdahl's law.
In terms of personpower, the Celeritas project has so far used 5 FTE-years spread over 2 years, and AdePT is similar.
Because this effort builds not just on \GEANTfour, but also on the lessons learned from earlier projects including Shift and \GEANTV, as well as the products of those efforts such as \VECGEOM, the total personpower involved is much greater.

Another avenue to achieve faster simulation is the use of machine learning to replace or augment existing ``classical'' (rule-based) simulation engines~\cite{Adelmann:2022ozp}.
Full replacement is most commonly pursued using generative models such as generative adversarial networks (GANs), variational autoencoders (VAEs), or normalizing flows (NFs).
Augmentation to improve lower-quality results, sometimes called refinement, can be applied to classical fast simulation engines or to generative models, using techniques similar to those above or regression-based approaches.
These approaches are largely still in development or just starting to be deployed~\cite{ATLAS:2021pzo}, so their ability to provide an acceptable level of physics fidelity at scale must be assessed carefully.
Beyond these ML algorithms, there are proposals to explore the application of differentiable programming directly to simulation engines~\cite{Adelmann:2022ozp,Dorigo:2022gqm}.
ML algorithms whose training and inference are performed using industry-standard frameworks are more easily portable to coprocessors such as GPUs, including at HPCs.
The possibility also exists to use coprocessors more cost-effectively and efficiently by performing the ML inference as a service, avoiding the need for each CPU to be equipped with a GPU~\cite{Harris:2022qtm}.

The above solutions, including GPU prototype transport engines and ML algorithms, are promising avenues for faster execution of detector simulation.
However, it is unlikely that these will completely supplant \GEANTfour, which provides broad support for myriad use cases.
In addition, the results of any new approaches are most often compared to \GEANTfour to prove their validity.
Therefore, some percentage of general purpose von Neumann CPU-based computing will be needed in HEP for the decades to come.
The extent to which CPU-based computing can be shifted to less general architectures and devices such as GPUs depends strongly on the availability of experts to develop and maintain the necessary software.
There has recently been a severe decrease in funding for permanent positions in HEP to support software developers who work on both these common tools and experiment-specific matters.
This threatens the persistence of important knowledge, which must be passed directly from person to person, and the stability and future of these crucial software packages that serve as a foundation for the entire field.
There is a strong and vocal consensus throughout the field that this trend must be reversed, with funding for technical positions not just restored, but increased further:
\begin{itemize}
\item ``...it is still necessary to support the existing code and make sure that sufficient funding and staffing is provided for maintenance and development of physics algorithms, as well as for adapting the code to any updated CPU hardware, operating systems and new compilers.''~\cite{HEPSoftwareFoundation:2020daq}
\item ``Unfortunately, the Bertini model [of hadronic interactions] has not been actively developed over the last few years due to the lack of personpower.''~\cite{Banerjee:2022}
\item ``A team of highly-skilled physicists and engineers is required to provide the necessary support and developments for Geant4, and the packages extending it, to meet the needs and challenges within the scope of the US HEP experimental program.''~\cite{Banerjee:2022jgv}
\item ``In addition, Geant4's common software, such as physics models, no longer receives any US maintenance funding... This is not sustainable for the long term viability of small experiments. Long term, experiment-agnostic Geant4 support is critical for the success of small experiments... It is therefore essential that there is funding for permanent software and computing experts.''~\cite{FASER:2022yqp}
\item ``However, U.S. funding for [the Geant4 toolkit] was discontinued in recent years... Continued
support for Geant4 is crucial to the design and construction of future experiments,
and for the interpretation of their results... The general challenge of maintaining software and computing talent is exacerbated in the direct detection community by the lack of long term, permanent positions within the experiments... It is therefore essential to provide funding for permanent software and computing experts.''~\cite{Kahn:2022kae}
\item ``DM and neutrino collaborations need scientists trained in data acquisition, simulation, and analysis at both the user and developer levels to achieve their science goals.''~\cite{Roberts:2022ezy}
\end{itemize}

\section{Physics Generators}\label{sec:evtgen}

The generator software landscape is complex and varied, in order to provide solutions to different aspects:
event (matrix element) generation, hadronization and parton shower modeling, underlying event tunes, matching/merging algorithms, multi-parton interactions, particle decays, cross section calculations, parton distribution functions, data formats, and analysis and reinterpretation.
There are many software packages that address some of these needs, which often include plugins to perform specific calculations and provide ways for users to plug in their own custom models and computations.
A rough and incomplete categorization includes:
\begin{itemize}
\item \begin{sloppypar}matrix element generators, which support leading-order (LO) as well as next-to-leading-order (NLO) and sometimes higher order calculations: \MGvATNLO, \POWHEG, \SHERPA, \ALPGEN, \WHIZARD, \AMEGIC, \COMIX, \HELAC;\end{sloppypar}
\item parton shower modelers, which also support LO event generation: \PYTHIA, \HERWIG, \SHERPA;
\item neutrino-nucleus interaction generators: \GENIE, \NEUT, \NUWRO, \GIBUU, \ACHILLES;
\item plugins for more specific tasks: \EVTGEN, \TAUOLA, \PHOTOS, \DIRE, \VINCIA;
\item higher-order differential cross section calculation: \MCFM, \MATRIX, \NNLOJET;
\item one-loop amplitude computation: \MADLOOP, \OPENLOOPS, \RECOLA;
\item utility software: \LHAPDF, \HEPMC;
\item analysis and tuning: \RIVET, \MCNNTUNES, \PROFESSOR, \APPRENTICE, \NUISANCE.
\end{itemize}
This wide array of software packages includes a corresponding variety of programming languages and styles---primarily Fortran, C++, and Python---as well as development and maintenance approaches.
Unlike detector simulation, in which \GEANTfour serves as a locus for numerous related efforts as described in Section~\ref{sec:detsim}, most of the software packages listed here are developed largely independently by separate groups.
The usage of common data formats and exchange specifications, such as LHEF, UFO, and \HEPMC, provides some amount of generic interoperability.

The percentage of WLCG computing time used for event generation at the CMS and ATLAS experiments is estimated to be 5--12\%, with similar or even more substantial requirements expected for ALICE and LHCb~\cite{HSFPhysicsEventGeneratorWG:2020gxw}.
However, these summary values hide large variances in tasks and generator software.
For example, \MGvATNLO has been found to be substantially faster than \SHERPA for generation of W+jets events with 0--2 extra partons, but much of the difference depends on parameter choices that may impact physical accuracy in some regions of phase space.
In general, for a given task, generator CPU usage scales linearly with the number of events generated.
However, moving from LO to NLO and higher orders, or increasing the multiplicity with additional partons~\cite{Hoche:2019flt}, for the same physics process typically causes a factorial increase in CPU usage, while different processes may still have very different baseline computational requirements.
The memory usage and multithreading abilities of different generator software are also highly variable and can lead to inefficiencies.
The range of programming languages and styles can make generators difficult to integrate directly in experiment software frameworks, another potential source of inefficiency.
The generator usage at neutrino frontier experiments has not been measured and profiled as systematically, and this subset of the field is generally smaller compared to the energy frontier.

Generator CPU usage is projected to increase to 8--20\% in the HL-LHC era, when more events and higher precision will frequently be required~\cite{CMSOfflineComputingResults,Collaboration:2802918}.
The increasing precision of upcoming neutrino experiments such as LBNF/DUNE and HyperK should be expected to require similar increases in experiment computing time devoted to event generation.
Further future experiments will have different generator needs depending on their initial states.
The different categories include high energy hadron colliders, lower (but still relatively high) energy lepton colliders, neutrino experiments, forward physics experiments (such as FASER and SND@LHC), transverse detectors (such as CODEX-b, AL3X, and MATHUSLA), the Electron-Ion Collider (EIC), and proposed muon colliders.
The details for the relevant tasks and generators in each category are vast and cannot be summarized here; Ref.~\cite{Campbell:2022qmc} presents a comprehensive treatment.
As a particular item of interest from the recent Snowmass effort, we highlight new techniques in development for muon collider physics: the use of the effective vector boson approximation for $2 \rightarrow n$ processes~\cite{Ruiz:2021tdt} and vector boson fusion as the dominant tool to study the electroweak sector at multiple TeV~\cite{Costantini:2020stv}.
These processes and approaches have been implemented in new versions of \MGvATNLO, sometimes requiring new features or more involved calculations to ensure stable and reliable results.
They illustrate the work that will be needed to model the physics at the next generations of HEP experiments.
The interdependence between lattice QCD, nuclear physics, and neutrino event generators is also important to highlight~\cite{Ruso:2022qes}.

The computational burden of an event generator depends strongly on the precision required, as described above; the calculation techniques used; and any other requirements placed on the output~\cite{HEPSoftwareFoundation:2020daq,HSFPhysicsEventGeneratorWG:2020gxw,Campbell:2022qmc}.
For example, the automated computation of NLO and higher order diagrams, as used in \MCATNLO, introduces events with negative weights, which can substantially reduce the statistical power of the final result.
There are also inefficiencies in phase space sampling; in slicing, biasing, or filtering to change kinematic features; and in merging to avoid double-counting between matrix element generators and parton shower modelers.
Further, experiments' use of generators may involve significant repetition of expensive calculations, especially to assess certain systematic uncertainties that cannot be represented as alternative event weights.
Recent efforts coordinated by the HEP Software Foundation (HSF) have identified and resolved several performance bottlenecks, some related to these issues~\cite{Campbell:2022qmc}.
There is ongoing work to continue to address these issues, especially negative weights, which can be reduced, but not yet eliminated, by certain new prescriptions that have not yet achieved wide deployment~\cite{HSFPhysicsEventGeneratorWG:2020gxw}.
In general, the development of new theoretical and mathematical tools to perform relevant computations has an outsized impact on event generator computing requirements.

At the time of the previous Snowmass reports, generator software was in some ways ahead of other subfields.
Machine learning algorithms were employed to derive PDFs~\cite{NNPDF:2014otw},
and usage of MPI for efficient utilization of existing CPU-based HPC systems was incorporated in \SHERPA, \ALPGEN, \MGvATNLO, and \MCFM, among others.
However, more recent initial efforts to port \MGvATNLO to use GPUs did not lead to a production quality result.
Revitalizing this task has been identified as a priority, and recent progress is promising.
An opportunity comes from the possibility to parallelize the matrix element calculations, which take up more than 90\% of the computing time for complex LHC physics processes.
Matrix element calculations are a perfect fit for CPU vectorization and hardware acceleration on GPUs.
The \MGGPU project, which uses custom CUDA kernels, shows speed improvement factors of ${\sim}1000$ on an Nvidia V100 GPU, as well as an improvement factor of 3 on CPUs by exploiting vectorization~\cite{Valassi:2021ljk}.
The \MADFLOW project, which uses a \TENSORFLOW backend, shows a speed improvement of more than an order of magnitude on an Nvidia Titan V GPU and an improvement factor of up to 3 on CPUs~\cite{Carrazza:2021gpx}.
A new approach called \BLOCKGEN for multi-gluon tree amplitudes shows similar improvements of roughly an order of magnitude on a V100 GPU and 2--6 on CPU, depending on the multiplicity~\cite{Bothmann:2021nch}.
Similar developments for other computationally-intensive generators such as \SHERPA also need to be prioritized and supported.
The use of ML for approximate matrix element calculations, including phase space integration and sampling as well as fully generative models similar to those discussed in Section~\ref{sec:detsim}, is also very promising, but still in relatively early R\&D and not yet ready for deployment.
Additionally, there are ongoing investigations into ML for parton shower modeling, hadronization, event (un)weighting, and tuning, among other applications~\cite{Campbell:2022qmc,Butter:2022rso}.

Beyond technical developments, other possible improvements that require coordination between different groups have been identified.
A common interface, which could also be described as a modular framework, would standardize how generators are used and plug into each other and other software~\cite{Ruso:2022qes,HSFPhysicsEventGeneratorWG:2021xti,Campbell:2022qmc}.
This would be useful not just for theoretical studies, but also for experiment use, especially small experiments with minimal personpower to implement, test, and maintain different interfaces.
It could also facilitate more collaboration among generator developers, for example to make faster progress on code modernization and related work.
The adoption of common data formats and standards, such as those provided by Les Houches, by more generators and communities could be encouraged as well.
While there are clearly a number of potential benefits, both coordination and personpower limitations would need to be overcome in order to realize them.
Sharing generator outputs, such as parton-level event records, is another idea to reduce both computation and disk usage~\cite{HSFPhysicsEventGeneratorWG:2020gxw}.
At the LHC, ATLAS and CMS often duplicate the work to generate the same physics process at the same precision and to store the result.
Avoiding this duplication would require coordination to standardize data formats, host the output, and ensure that the choices of model and generator parameters satisfy different analysis needs.
The existing LHC working groups that provide resources for cross sections and other theoretical calculations could serve as a model to organize this kind of coordination.
Data preservation organizations could act as a central hub for storage, and an increased emphasis on data preservation and access would also improve the field's capabilities to reinterpret and combine results~\cite{Campbell:2022qmc}.
More broadly, it has been suggested that a diverse and cross-cutting collaboration, inspired to MCNet in Europe with a long-term outlook, could bring together the U.S. event generator community to share resources and ideas~\cite{Campbell:2022qmc}.

The breadth and diversity of generator software brings many advantages and opportunities to HEP.
However, it also brings challenges, especially when it comes to keeping up with the increasing demands for both physics precision and computing efficiency.
Historically, there has not been a single obvious locus to coordinate common activities, even if MCNet and, more recently, HSF have provided useful forums, especially in the areas of training and computational efficiency, respectively.
Towards the future, more emphasis should be placed on establishing and coordinating common activities.
and improvements in this area should continue to be emphasized.
Projects to reduce the computational burden of event generation, for example by adapting to use GPUs, need a substantial increase in effort in order to be successful.
Some of the difficulties that generators face in computing are intrinsically tied to the nature of the calculations being performed, so fundamental research may be required to develop better methods.
On the whole, event generators are usually supported by small teams that frequently have little to no dedicated funding and cannot adequately plan for succession.
As with detector simulation, it is vital to the entire field that funding is provided for permanent positions to support, improve, and expand these crucial components of HEP software.
This is echoed throughout the white papers written during the Snowmass process:
\begin{itemize}
\item ``The physics Generators used in the field (eg. Pythia, GENIE, Madgraph, Sherpa) also suffer from lack of stable funding in a similar way.''~\cite{FASER:2022yqp}
\item ``The common needs to all direct dark matter detection experiments include: ... Continued support for event generators, including those developed as part of a national security program.''~\cite{Kahn:2022kae}
\item ``Participation in generator-related activities is poorly incentivized for both theorists and experimentalists, and opportunities to pursue neutrino generator development as one's primary research activity are rare... A need for greater coordination and prioritization of such activities is widely recognized in the neutrino scattering community. Despite promising initial discussions that have taken place in a series of recent workshops [68, 529--531]; however, neither a clear leadership structure nor significant institutional support to carry out the related work have yet emerged... The cooperation needed for success in neutrino generator development cannot occur to the extent that funding agencies impose rigid boundaries between [high energy and nuclear physics]... [Stitching together multiple models to achieve complete simulations] is ideally done in collaboration with theorists to minimize inconsistencies, but incentives for their direct involvement are currently poor.''~\cite{Ruso:2022qes}
\item ``[Theorists]... may not be motivated to work on software optimisations that are not ``theoretical'' enough to advance their careers. Generator support tasks in the experiments may also not be enough to secure jobs or funding for experimentalists pursuing a career in research.''~\cite{HEPSoftwareFoundation:2020daq}
\item Ref.~\cite{HSFPhysicsEventGeneratorWG:2021xti} specifically surveys the major collider event generators to report the expected level of support through the HL-LHC era: ``Funding and career perspectives are a major common issue reported by many generator experts... Generator support tasks in the experiments may also not be enough to secure jobs or funding for experimentalists pursuing a career in research. It is still not clear for people who are doing work on generators or related tools whether it is really possible to have a dedicated career path from now throughout HL-LHC. This is slowly being acknowledged from funding bodies, but the community has low expectations especially because there is no sign that hiring policies will be modified to provide reasonable funding for generator work. The delay caused by the lack of funding for generator work has already started to cause a loss of know-how in some generator packages.''
\begin{itemize}
\item \MGvATNLO: ``the collaboration are of the opinion, shared by other groups, that time to work on performance improvements, even by significant factors, is not well recognised by funding agencies... There are no dedicated funds for HL-LHC.''
\item \POWHEG: ``there are no funds specifically dedicated to the maintenance of the framework''
\item \SHERPA: ``Several of the Sherpa team are in permanent positions... In the UK, Sherpa posts at IPPP are seen as part of IPPP's core mission and therefore are expected to be treated as core in future funding reviews.''
\item \HERWIG: ``Funding is a serious issue... To fund the bread and butter of what the experiments need, one would need different specific funding.''
\item \PYTHIA: ``the continued development of Pythia8 throughout the HL-LHC era depends on funding agencies and laboratory priorities, though these are not expected to change. The biggest challenge to the collaboration is finding permanent positions for junior people so that they can continue their valuable contributions.''
\item \EVTGEN: ``The core EvtGen team has 4 members who have a fraction of their time funded for EvtGen through an UK STFC grant... There is some funding for a computer engineer to work on code redesign for multithreading.''
\end{itemize}
\item The developers of the \BEAGLE generator, which is used for inelastic scattering in nuclear collisions, state: ``The future development of this code is uncertain primarily due to lack of manpower and reliable funding... As discussed above, funding/manpower issues are currently limiting our ability to implement many of these improvements.''~\cite{Campbell:2022qmc}
\end{itemize}

\section{Continuum Field Theory Calculations}\label{sec:perturb}

Continuum field theory calculations include both precision high order perturbative calculations~\cite{Cordero:2022gsh} and conformal bootstrap theoretical calculations~\cite{https://doi.org/10.48550/arxiv.2203.08117}.

\subsection{Precision perturbative calculation}

Precision measurements at the LHC and future colliders require theory predictions with uncertainties at the percent level for many observables. Theory uncertainties due to the perturbative truncation are particularly relevant and must be reduced for a large variety of relevant processes.
At the high luminosity LHC, even rare processes like
Higgs production require theoretical control of cross sections at the 1\% level. 
Reference~\cite{Cordero:2022gsh} surveyed the theoretical community and received responses
covering 53 scientific publications. This Snowmass whitepaper gave clear motivation for precision as follows: 
\begin{itemize}
    \item ``To fully exploit the physics potential of current and future collider experiments, in particular to unambiguously identify signals of new physics, it is crucial to improve the precision of theoretical predictions.''
\end{itemize}
Since the 2013 Snowmass exercise, the state-of-the-art has moved dramatically and now next-to-next-to-next-to-leading order (\NcLO) QCD calculations for $2\to 1$ processes and \NNLO QCD calculation for $2\to 3$ processes and $2\to 2$ processes with internal masses are becoming state-of-the-art and reasonably fast on 
contemporary computing resources, while automation and more complex final states at \NNLO at \NcLO are goals for the Snowmass period. 

The computational challenge is to perform Feynman loop integrals within perturbative quantum
field theory, giving access to the scattering matrix and its analytic structures, often
involving non-standard special mathematical functions. The aim
is to, eventually, be able to numerically evaluate the expressions such that the error
is reliable, the result has sufficient precision, and the computation consumes acceptable resources.
Suitable analytical or symbolic preparation of the integrals and amplitudes must be performed
and the evaluation of these expressions during Monte Carlo integration over final state phase space, so that they may feed into event generators.

Analytical solutions of Feynman integrals in terms of special mathematical functions are preferred from an efficiency perspective, but often require process-specific knowledge or even new mathematical methods.
Numerical or semi-analytical methods can allow broader flexibility and better automation.
Numerical direct integration approaches are fully established and implemented in codes such as \PYSECDEC and \FIESTA, but are typically slow in evaluation times.
Novel semi-analytical methods based on series expansions seem promising and first public packages such as \DiffExp and \AMFlow became available recently. Such series expansion methods reduce evaluation times to the order of a minute per phase-space point for typical two-loop integrals.

QCD corrections to scattering amplitudes have been fully automated at NLO using libraries like
\HELACNLO, \MGvATNLO, \NLOX, \OPENLOOPS, and \RECOLA as well as many other public and private tools.
Two and higher loop calculations remain challenging but in the last five years significant progress has been made in many combinations of final state particle multiplicity and number of internal mass scales at two loops, and in more restrictive cases at three and four loops. Further the complete five loop
beta function has been calculated.

A core computational bottleneck for the calculation of both Feynman amplitudes and loop integrals is in the reduction of linear relations between Feynman integrals (integration-by-parts identities), with \FIRE, \REDUZE, \LITERED and \KIRA being example codes.
From a computational perspective the use of modular (finite field) arithmetic in the calculation
of integration by parts reduction is an interesting transformation of the computational requirement that may require radically different computing hardware solutions.
Reference~\cite{Cordero:2022gsh} notes significant difficulties fitting within the constraints of common HPC centers:
\begin{itemize}
    \item ``These novel techniques have opened the door to perform computer algebraic computations for quantum field theory on HPC systems. These novel methods typically use integer arithmetic rather than floating point arithmetic and, depending on the problem, they may require a significant amount of memory per core. The development of (public) codes for challenging amplitude calculations in a HPC friendly way is an ongoing effort. Currently, in some situations, one may need to resort to implementations that have large memory demands and require run-times that are hard to predict and possibly well beyond available batch limits on a given cluster.''
\end{itemize}
Given the importance of precision theoretical
calculation it is imperative that appropriate resources be available to enable them. 

\begin{SCfigure}[50][h]
    \caption{%
    \label{fig:multiloop}
    Computational costs of current state-of-the-art multi-loop calculations in the community, from Ref.~\cite{Cordero:2022gsh}. 
        }
    \includegraphics[width=0.5\textwidth]{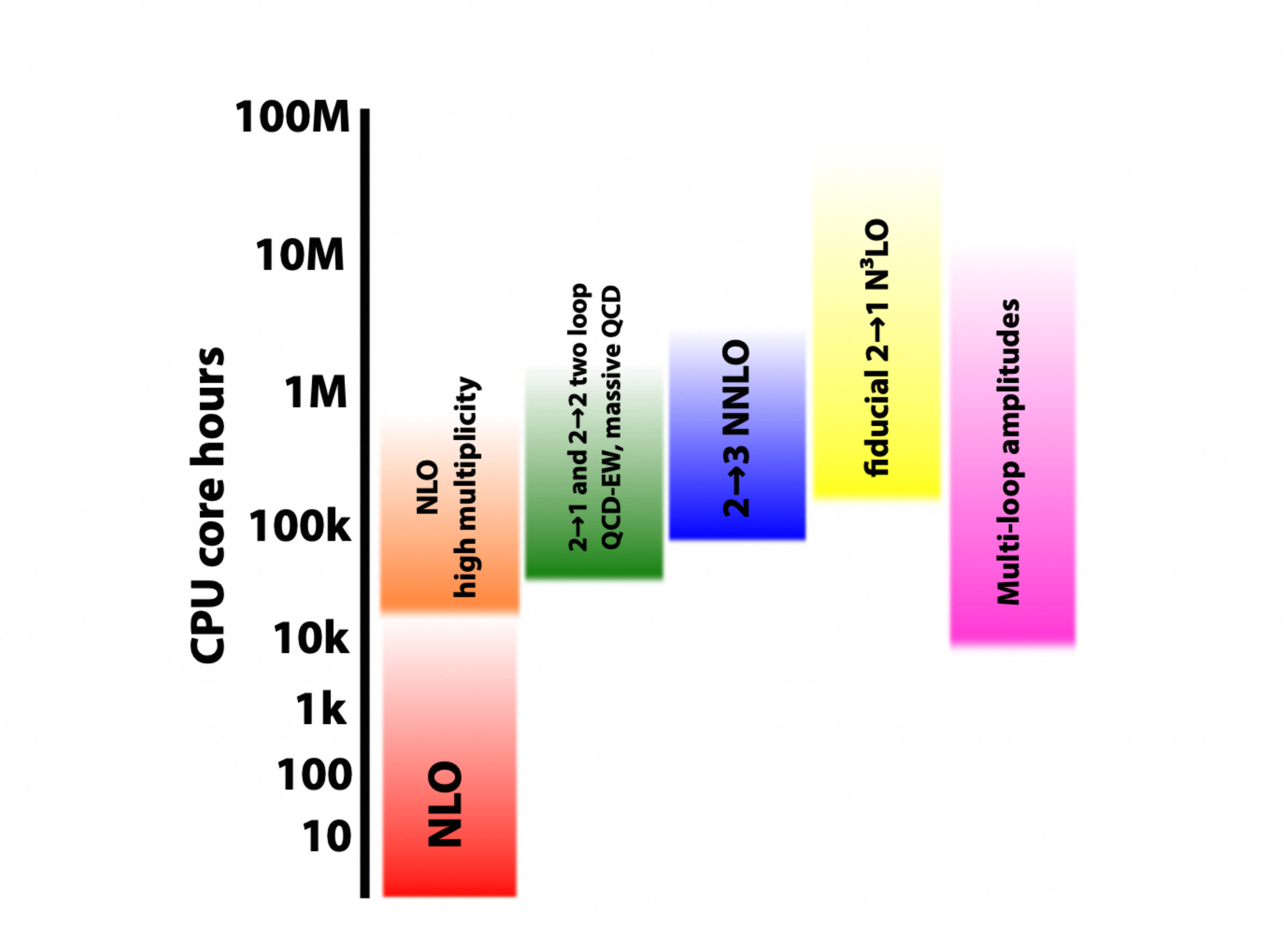} 
\end{SCfigure}

In 2013, a sample of then state-of-the-art
calculations took between 50,000 and 1M core hours. Since then, the per-core CPU performance has
improved by about a factor of 5 and the per-node performance by about a factor of 30. At the time, it was hoped to be able
to use efficiently either many-core architectures 
such as Xeon Phi or GPUs. 
The current state of the art has the additional complexity largely balanced by the growth in per-node performance, reaching up to about 10M core hours with current CPUs, as shown in 
Fig.~\ref{fig:multiloop}.
Parallelism is typically used within a node, 
and is indeed in part dictated by growth in core count relative to memory capacity. Multinode
parallelism is only exploited as throughput computing making use of trivial
parallelism by the majority of codes in the area.

Although there is some role for GPUs in numerical integration, they only match a modest part of the scientific requirement~\cite{Cordero:2022gsh}:
\begin{itemize}
    \item ``The use of GPUs is still unclear in our field, since many problems in our field rely on the numerical evaluation or handling of algebraic expressions which are large and/or require irregular memory access patterns. So far, GPUs have found application to cutting-edge problems with the numerical integration of sector decomposed loop integrals. A first step for future applications could include the efficient evaluation of one-loop amplitudes. This would help the huge computational requirements for NLO high-multiplicity evaluations, but also for the real emission integrations for \NNLO calculations. Since the efficient use of GPUs is still unclear, future computing for our community will still need to focus on providing CPU resources without attached GPU resources.''
\end{itemize}
Large memory and runtime limitations imposed by typical HPC environments mean that bespoke computing nodes and queues (wherever they are hosted) are required to enable these calculations.
Some of the job runtime constraints could be addressed with checkpointing virtual machines
but since complex software is involved, some even commercial like Mathematica, a dedicated
node environment may be the only practical solution. Machine learning does not at this time
play a large role, although some applications have been used in event generation.
The licensing of proprietary software like Mathematica or Maple also suggests dedicated
computing nodes for this purpose may be cost-effective.

\subsection{Conformal bootstrap}

The idea of the conformal bootstrap is to constrain and solve conformal field theories (CFTs) using physical consistency conditions like symmetry, unitarity, and causality. By relying on nonperturbative structures, bootstrap methods can work even in strongly-coupled systems where traditional perturbative techniques fail. Over the last few decades, the conformal bootstrap idea has crystallized into two concrete strategies: 1) exploiting exact solvability and 2) deriving bounds using sum rules with positivity constraints
and using convex optimization to extract information. When combined with numerical convex optimization, this leads to the numerical conformal bootstrap. The conformal bootstrap is distinguished by being an intrinsically non-perturbative method
that yields strict bounds on the theory.
Recent successes include precise determinations of critical exponents in physically-relevant theories such as the 3D Ising and O(N) models, constraints on theoretically important theories such as 4D N = 4 supersymmetric Yang-Mills theory and 6D $(2,0)$ SCFTs, and has inspired promising new ideas for bootstrapping S-matrices.
The numerical bootstrap is a rapidly-growing field and is anticipated to play a central role in CFTs over the next decade.

Targets include 3D scalar models,
$O(N)$ vector models in 3D,
and in the case of $N=1$ the 3D Ising CFT, while
the 3D O(2) model describes the
the superfluid transition in liquid helium as well as thin-film superconductors. Here the bootstrap
has helped address a controversy, where the results of experiment and the current best Monte Carlo simulations disagreed with each other with 8$\sigma$ significance,
giving results in remarkable agreement with those from Monte Carlo simulations.

A second target is 3D Fermionic Models, including
3D CFTs containing N Majorana fermions with Yukawa couplings to one or more scalar fields, often called Gross-Neveu-Yukawa model.
The simplest model with $N=1$ Majorana fermion coupled to a real scalar is believed to possess emergent supersymmetry and correspond to the minimal 3D $N=1$ supersymmetric extension of the Ising model, proposed to have a realization on the boundary of topological superconductors.
Preliminary studies applying the bootstrap to fermion 4-point functions showed the GNY models to be promising targets for the bootstrap, while further systems can be studied providing a supersymmetric systems including Wess Zumino models.

Progress in 3D and 4D gauge theories has been made including
including the conformal window of 4D QCD with 12 flavors, establishing a hint of contact with
lattice field theory near conformal window results and in $N=4$ supersymmetric Yang Mills theory.

From a computational perspective, the whitepaper identifies two key algorithmic challenges: i) to
find faster optimization methods, or ways to scale up current tools, and ii) to find efficient methods for exploring high-dimensional spaces of CFT data.
High precision arithmetic is required, and often not in
native hardware floating point, and used in bespoke semidefinite solver libraries such as SDPB.
SDPB can use distributed high precision linear algebra and
parallelize over many HPC nodes and hundreds to thousands of
cores. 

The authors are optimistic that further progress will:
\begin{itemize}
    \item ``enable a new round of fundamental discoveries in theoretical physics, including the numerical classification of CFTs with a small number of relevant operators, definitive answers to longstanding questions about conformal windows of gauge theories in 3 and 4 dimensions, more-or-less complete numerical solutions of the known maximally-supersymmetric CFTs (and their holographic duals), and new robust starting points for Hamiltonian truncation studies of gapped phases. Importantly, the numerical bootstrap can also be used as a discovery tool for finding previously unknown CFTs and as well as for identifying possible analytical solutions of known ones.''
\end{itemize}

The computational requirements so far have involve
the use of conventional cluster computers,
such as XSEDE and up to 32 nodes and 768 cores.
Calculations of different theories range from
200,000 core hours for the 3D Ising model to a few million core hours testing 100 to 1000 points in the CFT parametric space for the 3D $O(3)$ model. 
Large memory capacity is again significant requirement and is a barrier to efficiency, arising from the high precision (1000 bits) arithmetic and is required for accurately handling large cancellations in the calculation.
The authors identify significant software and computing hardware opportunities in future:
\begin{itemize}
    \item ``further scaling and improving this code, including implementing more sophisticated distributed linear algebra routines, improving memory management to allow scaling past ${\sim}500$ cores, finding ways to reduce precision requirements, and leveraging new hardware like GPUs and FPGAs. These engineering challenges are tightly coupled to physics: an increase in the scale of solvable semidefinite programs could allow one to explore larger systems of crossing equations and thereby access new CFTs.''
\end{itemize}

\subsection{Continuum field theory summary}
Both multi-loop perturbative calculations and numerical conformal bootstrap
have a similar computer hardware requirement. They both have a current
software reliance on CPU cores, and a large memory per core requirement.
Perturbative calculations have a sizable barrier to the use of GPUs,
while the conformal bootstrap approach, after significant software engineering,
 may be amenable to GPU or FPGA acceleration of high precision arithmetic.

\section{Lattice QCD}\label{sec:lattice}

Lattice gauge theory is a systematically improvable theoretical
tool for numerical evaluation of the Euclidean Feynman-path integral for
Quantum Chromodynamics (QCD).  Worldwide lattice gauge theory efforts 
directly support numerous high-energy 
physics experiments by calculating properties
of hadrons that are vital to interpretation of the experiments.
These efforts make significant use of supercomputers and are critically dependent 
upon continued computing advances.
The Flavour Lattice Averaging Group~\cite{Aoki:2021kgd} performs a critical
review of many important lattice QCD predictions every two to three years.
Snowmass Whitepapers have been contributed on the topics relevant to Lattice QCD \cite{https://doi.org/10.48550/arxiv.2202.07193,Ruso:2022qes,https://doi.org/10.48550/arxiv.2203.10998,https://doi.org/10.48550/arxiv.2204.07944,Boyle:2022ncb,Colangelo:2022jxc,https://doi.org/10.48550/arxiv.2203.12169,https://doi.org/10.48550/arxiv.2203.03230}.
USQCD also published in 2019 comprehensive set of 
topical whitepapers and submitted these
for consideration by Snowmass
\cite{Detmold:2019ghl,Bazavov:2019lgz,USQCD:2019hyg,Kronfeld:2019nfb,Cirigliano:2019jig,USQCD:2019hee,Joo2019}. 
The search for new physics requires a joint experimental and theoretical effort. Lattice QCD is an essential tool for obtaining precise model-free theoretical predictions of the hadronic processes underlying many key experimental searches.

The role of Lattice QCD in determining the
hadronic contributions to the anomalous magnetic moment of the muon was discussed in a Snowmass white paper~\cite{Colangelo:2022jxc}, by the international Muon $g-2$ Theory Initiative white paper~\cite{Aoyama:2020ynm} and in a white paper by the USQCD collaboration in 2019~\cite{USQCD:2019hyg}.
These calculations will be critical to the interpretation of results from the Fermilab Muon $g-2$ Experiment~\cite{Muong-2:2021ojo}, which is currently in excellent agreement with results from the earlier Brookhaven experiment~\cite{Muong-2:2006rrc}.

Lattice QCD also plays a key role in ab initio prediction
of nucleon structure and parton physics \cite{https://doi.org/10.48550/arxiv.2204.07944,https://doi.org/10.48550/arxiv.2202.07193}.
Understanding the neutrino-nucleus interaction is critical to the 
analysis of the Deep Underground Neutrino Experiment.
The interaction between an intermediate W or Z boson with 
with a quark confined inside a neutron or proton within the nucleus is theoretically
complex. This was discussed by USQCD in a 2019 white paper~\cite{Kronfeld:2019nfb}
and holistically the interplay between lattice QCD, EFTs, nuclear physics, phenomenology, and neutrino event generators was discussed in a contributed Snowmass Whitepaper~\cite{Ruso:2022qes}.

A comprehensive quark-flavor physics program
has been discussed in a USQCD whitepaper~\cite{USQCD:2019hyg}. 
Contributed Snowmass white papers include improving the
understanding of anomalies in B physics~\cite{OliverSnowmass}, 
emerging from the the Large Hadron Collider and studied further
at Belle II, as well as reconciling CP violation observed in kaon experiments with the standard model and exploring rare kaon decays~\cite{https://doi.org/10.48550/arxiv.2203.10998}.
These phenomena are typically highly suppressed in the standard model and therefore also offer promising avenues for the discovery of new physics.

As experimental measurements become more precise over the next decade, lattice QCD will play an increasing role in providing the needed matching theoretical precision. Achieving the needed precision requires simulations with lattices with substantially increased resolution.  With finer lattice spacing comes an array of new challenges. They include algorithmic and software-engineering challenges, challenges in computer technology and design, and challenges in maintaining the necessary human resources.  

The simulations at finer lattice spacing and larger volumes required to realize these goals introduce new challenges~\cite{Boyle:2022ncb}:
\begin{itemize}
    \item ``Meeting them requires new algorithmic research, novel computer hardware design beyond the exascale, improved software engineering, and attention to maintaining human resources.''
\end{itemize}

As a computational problem, Lattice gauge theory is performed on structured
Cartesian grids with a high degree of regularity and natural data 
parallelism. The approach formulates the Feynman path integral for QCD as a statistical
mechanical sampling of the related Euclidean space path integral. The sampling is performed
by Markov chain Monte Carlo (MCMC) sampling, using forms of the hybrid Monte Carlo (HMC) algorithm~\cite{Duane:1987de,Clark:2006fx,Luscher:2004pav}. 
Present algorithms for both MCMC sampling and Dirac solvers display growing limitations as substantially greater ranges of energy scales are included in our problem, an algorithmic challenge called critical slowing down.
The development of numerical algorithms is a significant intellectual activity that spans physics, mathematics, and computer science.

The central repeated operation is the solution of the
gauge covariant Dirac equation. 
The solution is usually performed using iterative Krylov solvers,
Newer multigrid~\cite{Brannick:2007ue,Luscher:2007se,Babich:2010qb} solvers
have demonstrated order-of-magnitude gains for Wilson fermions by
approximately and repeatedly handling degrees of freedom in the low
lying eigenspace as a form of preconditioner.
The US HEP program is presently focused on the
domain-wall and staggered approaches,
but corresponding gains for staggered and domain-wall-fermion discretizations
are a critical open research activity. 
A second algorithmic direction is the critical slowing down of
MCMC algorithms. 
These are being studied under Exascale Computing and
SciDAC projects and such support is critical to 
progress in the field.

Recent algorithmic directions with different ways of attacking the
same problem include applications of machine learning
to configuration sampling. This possibility has
been discussed in a dedicated Snowmass white paper~\cite{Boyda:2022nmh} and
will be covered in the report of topical group
CompF3 Machine Learning\cite{CompF3}.
Related Snowmass submissions on tensor networks
and quantum simulation will be covered in
the report of CompF6 Quantum Computing \cite{https://doi.org/10.48550/arxiv.2204.03381,https://doi.org/10.48550/arxiv.2203.04902}.

The Lattice QCD workflow is divided into two phases. First, a MCMC sampling phase generates
an ensemble of the most likely gluon field configurations distributed
according to the QCD action.
The ensemble generation is serially dependent and represents a strong scaling
computational problem. Ideally one would be able to use efficiently $O(10^4)$ computing nodes
on $O(256^4)$ data points. On the largest scales this becomes a halo-exchange
communication problem with a very large interconnect bandwidth requirement 
since the local data bandwidths vastly exceed those of 
inter-node communication.
In the second phase hadronic observables are calculated
on each sampled configuration where many thousands of quark propagators
are calculated and assembled into hadronic correlation functions.
This both allows more scope for amortizing
the setup cost of advanced algorithms like multigrid or deflation, and also has a high
degree of trivial parallelism. 
Scaling Monte Carlo sampling to many computational
nodes requires strong scaling and high
interconnect bandwidth. Scaling the hadronic observable calculations
is not as challenging as multiple configurations can be
analyzed at the same time, each job running on a moderate number of nodes.

A number of specific simulations, Table~\ref{tab:costs},
have been proposed with estimated costs
in a Snowmass white paper~\cite{https://doi.org/10.48550/arxiv.2203.10998},
and the same methodology can be used to estimate
the requirements of the ideal ensemble for flavor physics.
The final entry
is associated with
physics in the B-meson system indicated
in Snowmass white paper~\cite{OliverSnowmass}.
 A $256^3\times 512$ lattice at a lattice spacing $a = 0.04$ fm ($a^{-1} \sim 5$ GeV) would allow us to simulate up/down, strange, charm, and bottom quarks at their physical mass in
a 10 fm box with $m_\pi L = 7$ and requires more than a sustained Exaflop year. Several such calculations would be sought by the whole community.
These are a modest subset of the many many proposed calculations across
the Snowmass contributions, but set the scale of computing sources required in the area of Lattice QCD simulations in support of experiment at beyond-exascale.
Between flavor physics, nucleon structure, neutrino
scattering, and beyond-the-standard model physics, one could project four or five times
this requirement in aggregate across the entire field in the USA alone.
On this basis, Ref.~\cite{Boyle:2022ncb} estimated that:
\begin{itemize}
\item ``These simulation goals therefore clearly demonstrate a need
for computers at least 10x more capable than the coming Exaflop
computers during the Snowmass period. Since the performance is required to be delivered
on a real-code performance basis...
more than an order of magnitude improvement, perhaps, from both
algorithms and computing are required.''
\end{itemize}

\begin{table}[hbt]
\begin{center}
\begin{tabular}{|c|c|c|}
\hline
Lattice volume & $a^{-1}$ [GeV] & Exaflop hours\\
\hline
$32^3\times 64$	&1.4&	1.5\\
$40^3\times 96$	&1.7&	3.5\\
$48^3\times 64$	&2.1&	7.5\\
$48^3\times 96$	&1.8&	7.54\\
$64^3\times 128$&	2.4&	25\\
$96^3\times 192$&	2.8&	120\\
$64^3\times 256$&	2.4&	50\\
$96^3\times 384$&	2.8&	250\\
\hline
$128^3\times 512$& 5.0	&	12000\\
\hline
\end{tabular}
\end{center}
\caption{\label{tab:costs}
Proposed lattice volumes and cost estimates 
in \emph{sustained} Exaflop hours, scaled from current
simulations on Cori (NERSC) and Summit (ORNL).
Volumes and estimates are proposed in Snowmass white paper~\cite{https://doi.org/10.48550/arxiv.2203.10998}, while the final 
entry uses the same methodology to estimate the cost
of the most expensive proposed B-physics capable simulation.
}
\end{table}

The massive vector parallelism of lattice gauge theory is, 
in principle, amenable to GPU and possibly other acceleration.
This imposes a significant additional programmer overhead and the
most commonly used packages receive sufficient 
investment to use the complete
range of modern accelerated supercomputers and many of the largest
projects use allocations on DOE supercomputer resources.

Commonly used Lattice QCD software has been supported
by the DOE SciDAC and Exascale Computing Projects.
These include \GRID~\cite{Boyle:2022nef,Boyle:2016lbp,Boyle:2017gzg}, \MILC~\cite{DeTar:2018pyj,Gottlieb:2000tn}, \CPS~\cite{Jung:2014ata}, \Chroma~\cite{Edwards:2004sx} and \QUDA~\cite{Clark:2009wm}.
This has enabled major packages to support the most advanced GPU accelerated HPC computers using software interfaces such as
HIP, SYCL, and CUDA APIs in addition to giving good performance
on several CPU SIMD architectures. 
Newer interfaces like OpenMP 5.0
offload and C++17 parallel STL are planned to
be adopted as and when appropriate.

For smaller projects, especially where rapid development and programmer
productivity are at a premium, it is better and more cost effective
in terms of human effort to maintain access to a range of CPU resources.
USQCD institutional clusters at Brookhaven National Laboratory, Fermilab, and Jefferson Laboratory have
been instrumental in supporting the significant number 
of smaller and experimental projects that would not
achieve the return on investment to justify bespoke software development for multiple architectures.

Reference~\cite{Boyle:2022ncb} notes the heavy reliance on high performance computing places a particularly large dependence on highly tuned 
bespoke software in this field:
\begin{itemize}
\item ``One hopes that this might consolidate the
proliferation of programming interfaces as the lack of standardization
imposes a \emph{very significant software development burden} on the science community with
duplication of effort for multiple systems. This software development
underpins the entire community effort and requires support to make the Snowmass science goals feasible.''
\item ``The challenges exist on multiple
fronts: intellectual in developing algorithms that evade critical slowing down, software engineering to develop well-performing and portable code on an evolving range of supercomputers and programming models, and technical to remain engaged with the DOE HPC community as systems are planned and developed.''
\item ``The community activity is dependent on the existence of
bespoke software environments that enable
efficient simulation on rapidly evolving supercomputers,
which requires significant expertise to develop and sustain.''
\end{itemize}

USQCD has had senior staff, postdoctoral researchers, and post-graduate students funded under the Exascale Computing Project (ECP) and the SciDAC program.
 It has also funded the development of high performance software portable to Exascale hardware~\cite{Boyle:2022nef,Boyle:2017gzg}.
 It was noted as important that flexible high-performance software continues to be developed for a diverse range of architectures that tracks the DOE computing program:  
 \begin{itemize} 
\item ``the productivity of the community depends on large code bases... which do not have a secure model for development and support which places investments at risk.''
\end{itemize}

Much of the bespoke software development is performed in US National Laboratories. This is required to address HPC architectures as they emerge in the Computational Frontier, developing performance-portable high-level libraries to enable a write once and run anywhere approach that many domain scientists, both in laboratories and universities, can modify effectively.
This effort must track the evolution of computing.

Finally, it was noted that 
theoretical particle physics has been one of
the last area of physics to recognize the importance of
computation in forefront research~\cite{Boyle:2022ncb}. Continued effort is
urgently required to overcome this historical bias and
create a vibrant pool of skilled young faculty, and around 
them their Ph.D. students and research groups. 
Given the role of lattice gauge theory in confronting hadronic experiment, 
this field would particularly benefit from a program of joint Lab-University appointments.

\section{Conclusions and Recommendations}\label{sec:concl}

For scientific activities that consume a large amount of computing time and map
well to emerging HPC architectures, there is a wonderful opportunity to apply
enormous amounts of computation to solve incredibly complex problems.

However, common cross-cutting issues have been raised across many different scientific topics.
These issues are largely centered on the critical dependency between modern science and
computation and software. Computer architectures have necessarily become increasingly
complex, diverse, and challenging to program. 

In order to turn the potential of DOE computing into scientific deliverables, it is critical to develop skills in the community and to fund the required software and algorithmic development. Support for widely used scientific 
software packages is required to enable the most effective return.

Further, accelerated architectures
are not fully general purpose and present challenges to algorithms that are strongly
serial without any intrinsic vectorization or parallelism potential.
Even for those algorithms for which it is possible in principle
to use this hardware, the investment to develop software that runs efficiently
across many different platforms and programming environments requires a real
investment by the domain scientists. This investment of effort may not
always be a good value proposition.

For these two reasons, there is a continuing need for a mix of computing with both
accelerated and general purpose architectures.
This mixture should be sized appropriately to accelerate
 large scale computation where possible but also to effectively support science whose
 software cannot exploit acceleration.
 This may occur if it is not possible to port the underlying algorithms to compute accelerators or if a cost-benefit analysis identifies such a port as not being a cost-effective proposition.
 The latter case likely comprises a ``tail'' of many smaller computing tasks that run diverse software with limited development effort available.

Our recommendations on computing hardware are as follows:
\begin{itemize}
    \item {\bf Faster computer hardware}: For lattice gauge theory applications, there is a need for HPC resources ten times or more faster than planned Exascale computing systems.
    \item {\bf Support of right-sized CPU clusters}: There should be a right-sized (as large as required, as small as possible) provisioning
    of general purpose computational cores with high performance
    memory for important algorithms that do not easily map to computational accelerators. National Laboratory institutional (non-HPC) clusters contribute to this role. NERSC is a welcome example of providing both CPU and GPU/accelerator resources to handle a wider variety of problems relevant to HEP. These should accommodate a diverse range of requirements for software, large memory, and long running single node jobs.
    \item {\bf Universal programming interface}: The proliferation of programming interfaces is a barrier
    to portability and return on investment in software. Simplification and consolidation of the interfaces will improve scientific output.
    \item {\bf Software portability}: The difficulty of programming highly accelerated hardware with performance portability is significant. Appropriate support for software development and maintenance is essential to ensure
    success of much of the science discussed in this report.
    \item {\bf Automation of memory hierarchy}: Hierarchical memory is a significant burden on scientific 
    programmer productivity. Virtual memory paging systems can alleviate this and can be made efficient by using larger page sizes, if necessary. This may need substantial investment by computer vendors with robust encouragement from the Department of Energy.
    \item {\bf Early access to new computing hardware}: A key element of managing the science program is the early engagement with DOE HPC laboratory sites during the development, and years prior to installation, of major new facilities.  The lead time for porting to new architectures lies in the region of multiple years, and early engagement with emerging architectures is required to ensure timely scientific exploitation. 
\end{itemize}

The scientific software ecosystem should be nurtured to enable the proposed Snowmass
science. There are several steps that could be proactively taken to ensure
the maximum scientific return is delivered.

\begin{itemize}
\item {\bf Hardware accelerator-friendly, portable common tools}: Specific projects to adapt common software packages to run efficiently on GPUs, including those at new HPCs, have shown initial promise with more than order of magnitude speed improvements.
These projects should be continued with sufficient effort to deliver complete products and to incorporate portability solutions to support different coprocessor devices and architectures.
Adoption of this new software will additionally require expertise devoted to integration and usage in experimental software frameworks, analysis tools, etc., which should be supported as discussed below.
The exploration of new machine learning-based methods, as well as technical improvements to existing (CPU-based) software, should also continue.
\item{\bf Best practices for common software}:
Whenever possible, interoperability and common data structures should be encouraged.
Existing standard libraries and underlying software technologies should be used when available and feasible.
Interoperability, portability, and common, shared software are fundamental pillars of international scientific collaboration and will become increasingly important in the future.
\item{\bf Lab-supported software development}:
Just as the large experiments require talented permanent staff
at the Labs to engineer experiments, theoretical calculation and
simulation in HEP experiment, in high energy theory, in cosmology, and in large particle accelerator building have large communities often dependent on sophisticated
long-term software systems.
Career paths must be created to retain
some of the most talented experts in software and algorithms.
A larger permanent laboratory staff of software development experts could help address HPC architectures as they increase in importance within the Computational Frontier.
\item {\bf Organization for long-term common software maintenance}:
Despite the numerous successes and broad adoption of common software tools across all topical areas in this report, the future of these projects is in jeopardy because of an extreme lack of funding.
One priority should be the development, maintenance, and support for common software tools in the areas of accelerator modeling, detector simulation, and physics generators.
Because this kind of long-term effort to sustain cross-cutting software does not fit easily into existing modes of funding, new processes and organizations should be established in cooperation with funding agencies.
\item {\bf Continuous collaboration with ASCR}: A lot of progress has been made in high energy physics applications through collaborations with programs supported by ASCR under the SciDAC and ECP projects. We believe that
such collaborations should be encouraged and continuously supported for future computational modeling in high energy physics. The community found the application development
focus of the ECP project particularly effective. 
\item {\bf Joint lab-university tenure track appointments}:
 We believe the DOE should 
seek to foster the continued development of intellectual leaders
in theoretical calculation and simulation. The health of the field 
requires a similar
cohort of individuals at the best universities, reflecting the
intellectual vigor and potential of this area to
contribute to DOE scientific goals. The creation of
such positions can be stimulated by DOE-funded joint, five-year, tenure-track appointments. 
A good example might be the Jefferson Lab University Relations program of joint and bridged faculty appointments in nuclear physics. 
\item {\bf Training}: Accessible training in both novel architectures and AI/ML with a low barrier to access for graduate students, postdoctoral researchers, and domain scientists is critical to ensuring the skills base exists
for productive science in theoretical calculation and simulation.
The use of cross-platform performance portable APIs should
be encouraged to maximize return on investment in software.
Hackathons and code examples are particularly useful.
\end{itemize}

\section*{Acknowledgments}

P. Boyle has been funded by the U.S. Department of Energy, Office of Science, Office of Nuclear Physics under the Contract No. DE-SC-0012704 (BNL).
K. Pedro is supported by the Fermi National Accelerator Laboratory, managed and operated by Fermi Research Alliance, LLC under Contract No. DE-AC02-07CH11359 with the U.S. Department of Energy.
J. Qiang is supported by the U.S. Department of Energy under Contract No. DE-AC02-05CH11231 (LBNL).

\bibliographystyle{JHEP}
\bibliography{Computation/CompF02/myreferences}







\end{document}